\documentstyle[twocolumn,aps,prl,psfig,floats]{revtex}

\begin{document}

\draft

\twocolumn[\hsize\textwidth\columnwidth\hsize\csname @twocolumnfalse\endcsname

\title{Weak ferromagnetism and field-induced spin reorientation in 
K$_2$V$_3$O$_8$}
\author{M.D. Lumsden$^1$, B.C. Sales$^1$, D. Mandrus$^{1,2}$, 
S.E. Nagler$^1$, and J.R. Thompson$^{1,2}$}
\address{$^1$ Oak Ridge National Laboratory, P.O. Box 2008, Oak Ridge, TN, 
37831 U.S.A.}
\address{$^2$ Department of Physics, University of Tennessee, Knoxville, TN,
37996-1200 U.S.A.}
\maketitle
\begin{abstract}
Magnetization and neutron diffraction measurements indicate long-range
antiferromagnetic ordering below $T_N$=4 K in the 2D, S=1/2 Heisenberg 
antiferromagnet K$_2$V$_3$O$_8$.  The ordered state exhibits
``weak ferromagnetism" and novel, field-induced spin reorientations.
These experimental observations are well described by a classical, 
two-spin Heisenberg model incorporating Dzyaloshinskii-Moriya interactions 
and an additional c-axis anisotropy. This additional anisotropy can be
accounted for by inclusion of the symmetric anisotropy term recently described
by Kaplan, Shekhtman, Entin-Wohlman, and Aharony.  This suggests that 
K$_2$V$_3$O$_8$ may be a very unique system where the $qualitative$ behavior
relies on the presence of this symmetric anisotropy.

\end{abstract}

]


One of the most exotic and intriguing anisotropies which can occur in
magnetic solid state systems is the antisymmetric
Dzyaloshinskii-Moriya (DM) interaction.  These interactions, originally
proposed by Dzyaloshinskii \cite{Dzyal} 
and later derived by Moriya \cite{Moriya2} through extension
of Anderson's superexchange model \cite{Anderson}, 
arise from spin-orbit coupling in
the presence of sufficiently low local lattice symmetry.  The form of
this interaction is typically written 
${\bf D} \cdot ({\bf S_1} \times {\bf S_2}$) where ${\bf D}$ is a vector
determined by the symmetry of the lattice.  Competition between the vector
product in the DM interaction and the scaler product in the Heisenberg
interaction often leads to novel phenomena such as spin-canting \cite{Moriya1}
and spiral spin structures \cite{Ishikawa1,Ishikawa2,Lebech,Zheludev}.

Relatively recently, Kaplan \cite{Kaplan}, and later
Shekhtman, Entin-Wohlman, and Aharony (KSEA) \cite{SEA1,SEA2,SEA3} 
(motivated by relevance to parent compounds of high-Tc superconductors), 
re-examined the DM interaction with particular emphasis on an
often ignored symmetric component. 
This symmetric anisotropy always accompanies the typical antisymmetric
DM term but is typically ignored as its strength is an order of magnitude
weaker.  KSEA showed that, in the 
general case where there is no $frustration$, this symmetric term cannot
be ignored as its inclusion restores the original O(3) symmetry of the 
Heisenberg Hamiltonian.  We report in this letter the novel properties 
of a S=1/2, 2d Heisenberg system, K$_2$V$_3$O$_8$, and suggest that 
they can be explained by a model including both symmetric
and antisymmetric DM interactions.  This implies that K$_2$V$_3$O$_8$ is a 
novel system where the $qualitative$ behavior relies 
heavily on inclusion of the symmetric component of the DM interaction.

K$_2$V$_3$O$_8$ crystallizes in a tetragonal unit cell with space group
P4bm and lattice constants a=8.870~\AA~ and c=5.215~\AA \cite{Galy}.  
The structure is shown in Fig. 1(a) and consists of magnetic V$^{4+}$-O$_5$ 
pyramids and non-magnetic V$^{5+}$-O$_4$ tetrahedra with interstitial 
K$^{+}$ ions.  Previous magnetic measurements consisted of powder magnetization
from 5-300 K \cite{Liu} which were best described 
by a 2D Heisenberg model with coupling constant J=12.6 K and a g-value of 1.89. 
EPR measurements on single crystals suggest very small anisotropy with 
g-values of g$_{c}$=1.922 and g$_{ab}$=1.972\cite{Pouchard}.  
Liu et al. \cite{Liu} suggested from observation of field-dependent 
magnetization that K$_2$V$_3$O$_8$ may order at lower temperatures.

Single crystal plates of K$_2$V$_3$O$_8$ 
(typical dimensions: 1$\times$1$\times$0.1 cm$^3$)  were
grown in a platinum crucible  by cooling appropriate  amounts  of VO$_2$ in a
molten KVO$_3$ flux. Additional  details of the crystal growth will be
reported elsewhere.\cite{Sales}  One of these single crystals
was mounted in a Quantum Design SQUID magnetometer equipped with a
sample rotator and the magnetization is shown in Fig. 2 as a function of both
temperature (a) and applied magnetic field (b) for fields
applied along both the c-axis and within the tetragonal 
basal plane.  The M(T) data (Fig. 2(a)) were measured by cooling in the
presence of an applied magnetic field while the M(H) data (Fig. 2(b)) 
were taken by cooling to base temperature in zero-field.
From Fig. 2(a), the magnetization is seen to be very isotropic for
temperatures in excess of about 8 K confirming the Heisenberg nature
of the interactions.  At high temperatures, we observe a Curie-Weiss 
susceptibility with an effective moment ($\mu$=g$\mu_B$[S(S+1)]$^{1/2}$) 
of 1.7$\mu_B$ per formula unit (FU).  This result is consistent with the 
presence of two non-magnetic V-ions per FU and one magnetic ion with S=1/2.

The temperature dependence in the presence of a weak
magnetic field shows a clear ordering phase transition below a temperature 
of about 4 K.  The rapid decrease of the magnetization 
below this temperature with field applied along the c-axis is indicative
of antiferromagnetic ordering state with an easy axis along c.
However, measurements within the basal plane show a clear ferromagnetic
enhancement upon entering the ordered state.  The 
ferromagnetic ordered moment is small ($\sim$10$^{-2}$ $\mu_B$) and thus we 
conclude that the system is antiferromagnetic with spins primarily
along the c-axis and a small canted moment in the basal plane.  This 
represents the first observation of an ordering phase transition 
in this material.  

The behavior of the magnetization as a function of field for temperatures
well below the N\'eel temperature is shown in Fig. 2(b).  For field applied
along c, we observe an abrupt increase in magnetization for fields
in excess of about 0.85 T consistent with a discontinuous spin-flop
phase transition, further confirming the easy nature of the 
c-axis.  For fields within the basal
plane, we observe an intriguing field-induced phase transition which
occurs at a magnetic field of H$_{SR}$=0.65 T.  There was no evidence of 
irreversibility in M(H) for either field orientation
or in angular rotations of the sample in the 
presence of an applied field and no evidence of thermal hysteresis was
observed.

\begin{figure}[bt]
\centerline{
\psfig{file=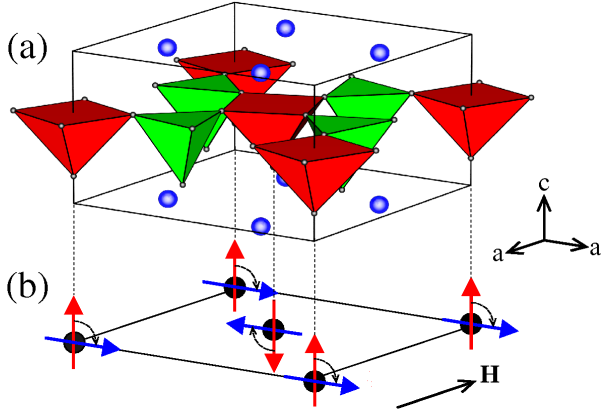,width=\columnwidth,clip=}}
\vspace{2ex plus 1ex minus 0.5ex}
\caption{
(a) The crystal structure of K$_2$V$_3$O$_8$ composed of
 magnetic V$^{4+}$-0$_5$ pyramids and non-magnetic V$^{5+}$-0$_4$ tetrahedra. 
(b) The projection of the
V$^{4+}$ positions showing the location of the magnetic moments.  The
red arrows represent the zero-field spin configuration and the blue
arrows denote the behavior of the system in the presence of a basal
plane magnetic field (in the direction shown by {\bf H}).
}
\label{fig1}
\end{figure}

To investigate the detailed microscopic 
spin arrangement, neutron diffraction measurements were performed on the
HB1A and HB1 triple axis spectrometers at the High Flux Isotope Reactor
in Oak Ridge National Laboratory.  For the measurements in zero-field, the 
sample was mounted in the ($h$0$l$) scattering plane and comparison of 
scattering at 1.6 K and 10 K indicated the presence of magnetic Bragg peaks 
(in the ($h$0$l$) plane) described by integer indices with $h$ odd.  These 
peaks suggest a magnetic unit cell in which the corner and face-center 
moments within the basal plane are aligned antiparallel with parallel 
alignment along the c-axis.  Examination of the integrated intensities of 
6 magnetic reflections indicated that the moments lie along the c-axis 
consistent with conclusions drawn from the magnetization measurements.  The 
value of the ordered moment ($\mu$=g$\mu_B$S) was found to be 0.72(4) $\mu_B$ 
at T=1.6 K, reduced from the expected value of 1 $\mu_B$. This can likely 
be attributed to quantum fluctuations 
in the low-dimensional S=1/2 system.  The zero-field magnetic structure
is shown schematically by the red arrows in Fig. 1(b).

To examine the behavior in the presence of a basal plane magnetic field, 
the crystal was mounted in the ($h$0$l$) scattering plane and a vertical 
field applied along the (010) direction.
The field dependence of the 
(100) and (101) magnetic Bragg peaks are shown in Figs. 3(a) and 3(b).
As the intensity of magnetic neutron scattering is 
proportional to the component of magnetic moment normal to the wavevector
transfer, Q, the observed disappearance of the (100) magnetic Bragg
peak and the concomitant 
enhancement of the (101) reflection are consistent with a 
continuous rotation of the spins from the (001) to the (100) direction
with increasing magnetic field strength.  The rotation is complete at a 
field strength of H$_{SR}$=0.65 T in excellent agreement with the anomaly
in the magnetization.  This is shown schematically by the blue arrows in
Fig. 1(b).  Thus, we conclude that this transition is a peculiar 
spin-reorientation phase transition whereby the spins rotate from the c-axis 
to the basal plane while remaining normal to the applied field direction.

\begin{figure}
\centerline{
\psfig{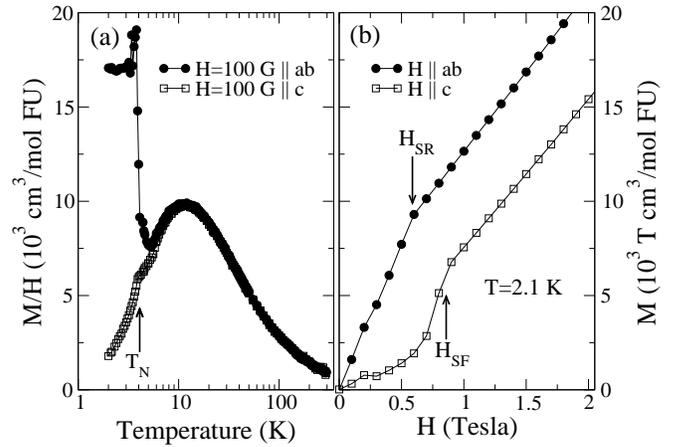}}
\vspace{2ex plus 1ex minus 0.5ex}
\caption{
(a) The temperature dependence of the DC susceptibility in a small
applied magnetic field.  For temperatures in excess of about 8 K, 
the system is seen to be very isotropic.  Below 4 K, the system
undergoes a phase transition to an antiferromagnetic state
characterized by an $easy$ c-axis and
weak ferromagnetic behavior in the basal plane.  (b) The magnetization
as a function of applied magnetic field for temperatures well below
the ordering temperature.  Field-induced phase transitions are observed
for field applied in the basal plane (H$_{SR}$) and 
along the c-axis (H$_{SF}$).
}
\label{fig2}
\end{figure}

A second crystal was mounted in the (hk0) scattering plane to allow 
measurements with field applied along the c-axis.  The
results of these measurements are shown in Fig 3(c) and (d).  
As a function of applied magnetic field, the (100) Bragg peak intensity
drops abruptly upon passing through a field of 0.85 T suggesting a 
discontinuous reorientation of the spins.  The temperature dependence
in an applied field of 1 T shows a strikingly different behavior for 
the (100) and (010) Bragg peaks and the weaker intensity for (010) suggests
that the moments have ``flopped" from the c-axis into the basal plane
selecting a specific direction within the basal plane in close
proximity to the (010) direction.  No evidence for a structural distortion
has been observed and the selection of a specific direction likely results  
from a slight misalignment of the magnetic field with the c-axis.
This misalignment was less than 1$^\circ$ for the measurements performed 
but even this level may induce a net field 
in the basal plane of sufficient strength to overcome domain energies thus
selecting a specific magnetic domain.  Such a domain selection has been observed
in the tetragonal antiferromagnet MnF$_2$ \cite{Felcher}.

\begin{figure}
\centerline{
\psfig{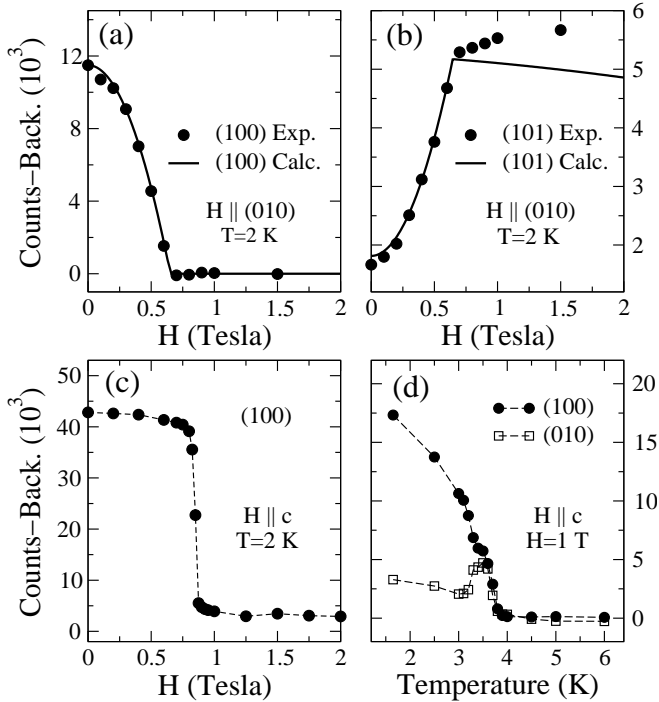}}
\vspace{2ex plus 1ex minus 0.5ex}
\caption{
The background subtracted intensity of diffracted neutrons as a function 
of applied magnetic field (along the (010) direction), for temperature well 
below the N\'eel temperature, is shown for the (100) and (101) magnetic Bragg 
peaks in panels (a) and (b).  The
solid lines represent calculations based on the model described in the
text. (c) The field dependence of the (100) Bragg reflection with 
field applied along the c-axis, for temperature 
well below the ordering temperature, showing a discontinuous spin-flop
phase transition.  (d) The temperature dependence of 
(100) and (010) in an applied field of 1 T.
}
\label{fig3}
\end{figure}

As mentioned in the introduction, weak ferromagnets often have at 
their origin the competition between the antisymmetric DM interaction
and the symmetric Heisenberg interaction.  To consider the DM interactions
in K$_2$V$_3$O$_8$, we will consider the projection of the crystal structure
onto the tetragonal basal plane as shown in Fig. 4(a).  The plane of inversion
symmetry between near-neighbor V$^{4+}$ spins results in DM vectors
located in that inversion plane \cite{Moriya1}
giving components along (110) and (001) directions,
denoted by ${\bf D_{xy}}$ and ${\bf D_z}$ respectively (these directional
vectors are indicted in the figure relative to the central site).  The 
four-fold rotation symmetry through the central site gives rise to the
configuration of ${\bf D}$ vectors shown in the figure.  From the
neutron diffraction measurements, we know that the magnetic unit cell 
contains two spins in both zero and non-zero fields and, hence, we will
consider a two-spin, classical Hamiltonian.  Noting that in averaging over
the the near-neighbor spins within a unit cell, 
the components of the DM vectors in 
the basal plane (i.e. the ${\bf D_{xy}}$ terms) cancel,   
we write the ground state energy as
\begin{equation}
 F=J {\bf M_1} \cdot {\bf M_2} + D_z [{\bf M_1} \times {\bf M_2}]_z
\end{equation}
where ${\bf M_1}$ and ${\bf M_2}$ are the sublattice magnetizations
and $[{\bf M_1} \times {\bf M_2}]_z$ is the z-component of the vector 
cross-product.

\begin{figure}
\centerline{
\psfig{file=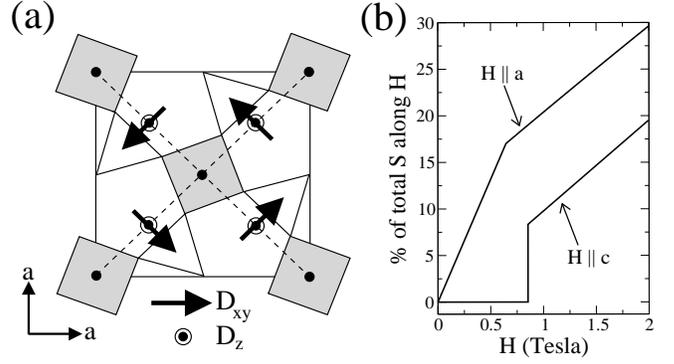,width=\columnwidth,clip=}}
\vspace{2ex plus 1ex minus 0.5ex}
\caption{
(a) The projection of the crystal structure onto the tetragonal basal
plane.  The magnetic moments are located at the center of the shaded 
squares and the triangles represent the V$^{5+}$-O$_5$ tetrahedra.  The
components of the Dzyaloshinskii-Moriya vectors (D$_{xy}$ and D$_z$)
are shown in the 
figure relative to the central spin.  (b) The calculated component of 
magnetic moment along the field direction as a function of applied field
for field applied along both
a and c.  This result compares very well to the M(H) data plotted
in Fig. 2(b). 
}
\label{fig4}
\end{figure}

This ground state energy
will be minimized when the spins are in the basal plane and, consequently,
an additional c-axis anisotropy is required to account
for the easy c-axis observed experimentally. 
Thus, we add a c-axis 
anisotropy of the form $E_zM_{1z}M_{2z}$ and, with addition of the 
Zeeman energy for nonzero field, we arrive at the ground state energy,
\begin{eqnarray}
F &=& J {\bf M_1} \cdot {\bf M_2} + D_z [{\bf M_1} \times {\bf M_2}]_z + 
E_zM_{1z}M_{2z} \nonumber\\
&+& g\mu_B{\bf H} \cdot [{\bf M_1} + {\bf M_2}].
\end{eqnarray}
This rather simple, classical ground state energy describes all the 
low temperature properties of the system.  For appropriate values of $D_z$
and $E_z$, the spins will point along the c-axis with canting observed
in small fields due to
the normal competition between DM and Heisenberg interactions.  
For fields applied in the basal plane, the Zeeman energy and
the DM interaction favor arrangement of the spins in the basal plane.
This competes with the c-axis anisotropy causing a continuous rotation
of the spins to the basal plane with increasing field strength.  
The spins remain primarily normal to the 
applied field direction to minimize the Zeeman energy.  Finally, with 
field applied
along the c-axis, a spin-flop transition occurs due to competition
between Zeeman energy and the anisotropies in the system causing 
an abrupt rotation of the spins into the basal plane.  

As we have two field-induced phase transitions, we can estimate the values for
D$_z$/J and E$_z$/J required
to produce phase transitions at the measured field values (using J=12.6K 
and g=1.89 as determined from Liu et al. \cite{Liu}).  This results in 
a value of D$_z$/J=0.22 and E$_z$/J=3.8$\times$10$^{-2}$
(this value of D/J seems large considering the 
measured g-values \cite{Liu,Pouchard} but 
compares favorably to the DM spiral system BaCuGe$_2$O$_7$
which has a similar basal plane structure and for which D/J was
found to be $\sim$0.18 \cite{Zheludev}). Using
these values, we can calculate the induced magnetic moment along 
the field direction for field applied along both the a- and c-axes and the 
results are plotted in Fig. 4(d) as a function of applied magnetic field in 
Tesla.  These results can be directly compared to the M(H) data plotted in 
Fig. 2(b) and the qualitative agreement is remarkable suggesting that this 
relatively simple model adequately describes the low temperature
properties of the system.  To further emphasize
this, we have superimposed on Fig. 3(a) and 3(b) the calculated magnetic
Bragg peak intensity for the (100) and (101) reflections in the presence
of an applied magnetic field along the (010) direction.  These results
are represented by the solid lines in Figs. 3(a) and 3(b) and the agreement 
between the calculations and the data is excellent, particularly
for the (100) reflection.  The agreement 
for the (101) reflection is very good for field strengths up to H$_{SR}$ but
seems to deviate for higher fields.  The reason for this disagreement
is unclear but could result from a field-dependent $<$S$>$.  Such a scenario 
can occur via field-dependent suppression of quantum fluctuations, as 
has been observed in some ABX$_3$ antiferromagnets \cite{Borovik}.

To shed some light on the nature of the c-axis anisotropy, we calculate
the symmetric anisotropy term of KSEA.  
The addition of this term causes a rescaling
of J to (J-D$^2$/4J) and the inclusion of a symmetric anisotropy term 
(1/2J)${\bf M_1}$$\cdot$${\bf A}$$\cdot$${\bf M_2}$ 
where ${\bf A}$ is a 3$\times$3
matrix with elements A$_{uv}$=D$_u$D$_v$ \cite{SEA2,Yildirim}.  
Making these changes to the Hamiltonian and averaging over near-neighbor 
spins produces the ground state energy,
\begin{eqnarray}
F&=&(J-{{D_z^2} \over {4J}}) {\bf M_1} \cdot {\bf M_2} + 
D_z [{\bf M_1} \times {\bf M_2}]_z \nonumber\\
&+& ({{D_z^2} \over {2J}}-{{D_{xy}^2} \over {4J}})M_{1z}M_{2z}+
g\mu_B{\bf H} \cdot [{\bf M_1} + {\bf M_2}].
\end{eqnarray}
This energy has precisely the same form as Eq. 2 and, consequently, the c-axis
anisotropy falls out naturally upon inclusion of both the
symmetric and antisymmetric anisotropies.
The maximum value for the c-axis anisotropy in Eq. 3
is D$_z^2$/2J which, for D$_z$/J of 0.22, gives a maximum anisotropy 
(E$_z$/J) of 2.4$\times$10$^{-2}$.  Within the level of approximation
involved in this model, this is in reasonably good agreement
with the above value of 3.8$\times$10$^{-2}$.  Consequently, we conclude
that K$_2$V$_3$O$_8$ represents
a unique system where the $qualitative$ features rely
heavily on the inclusion of the KSEA symmetric anisotropy term.  
The only known, clear evidence of this symmetric anisotropy occurs in 
the DM spiral system Ba$_2$CuGe$_2$O$_7$ where the addition of this term
is necessary to produce $quantitative$ agreement between experiment and
theory \cite{Zheludev2}.  As a caveat, it is important to note that we 
cannot rule out other possible mechanisms for this additional anisotropy.  
Rather, we simply note that inclusion of this symmetric anisotropy alone, 
which is necessary to properly consider DM interactions, seems to adequately 
describe the properties of the system.

In summary, we have observed an ordering phase
transition in K$_2$V$_3$O$_8$ with a 
primarily antiferromagnetic ordered state accompanied by weak
ferromagnetism.  Field-induced spin reorientation phase 
transitions have been observed with field applied along both the
c-axis and in the tetragonal basal plane.  The spin arrangement
has been determined by single-crystal neutron diffraction measurements.
Remarkably, this rich magnetic behavior can be well described by a simple,
classical, two-spin Heisenberg model with inclusion of DM interactions
and an additional c-axis anisotropy.  This additional anisotropy can,
at least partially, be accounted for by inclusion of the KSEA symmetric
anisotropy term making K$_2$V$_3$O$_8$ a unique system where introduction
of this interaction is necessary to describe the $qualitative$ behavior
of the system.

We would like to acknowledge stimulating discussions with G. Murthy,
M.W. Meisel, R. Coldea, and D.A. Tennant.
Oak Ridge National Laboratory is managed by
UT-Battelle, LLC, for the U.S. Dept. of Energy under 
contract DE-AC05-00OR22725.


%
%


\end{document}